\documentclass[12pt]{article} 
\usepackage{graphicx,subfigure,latexsym}

\newcommand{\be}{\begin{equation}}
\newcommand{\ee}{\end{equation}}
\newcommand{\bear}{\begin{eqnarray}}
\newcommand{\eear}{\end{eqnarray}}
\newcommand{\ba}{\begin{array}}
\newcommand{\ea}{\end{array}}

\begin{document}

\begin{titlepage}
\vfill
\begin{flushright}
{\normalsize arXiv:xxxx.xxxx[hep-th]}\\
\end{flushright}

\vfill
\begin{center}
{\Large\bf Holographic two dimensional QCD and Chern-Simons term  }

\vskip 0.3in

Ho-Ung Yee\footnote{e-mail:
{\tt hyee@tonic.physics.sunysb.edu}}
and
Ismail Zahed\footnote{e-mail:
{\tt zahed@tonic.physics.sunysb.edu}}
\vskip 0.15in
 {\it Department of Physics and Astronomy, SUNY Stony Brook,} \\
{\it Stony Brook, NY 11794-3800 }\\[0.3in]

{\normalsize  2011}

\end{center}

\vfill

\begin{abstract}
We present a holographic realization of large $N_c$ massless QCD in two dimensions using a
D2/D8 brane construction.  The flavor axial anomaly is dual to a three dimensional Chern-Simons
term which turns out to be of leading order, and it affects the meson spectrum and holographic renormalization in crucial ways.
The massless flavor bosons that exist in the spectrum are found to decouple from the heavier mesons, in agreement with the general
lore of non-Abelian bosonization.  We also show that an external dynamical photon acquires a mass through 
the three dimensional Chern-Simons term as expected from the Schwinger mechanism. Massless two
dimensional QCD at large $N_c$ exhibits anti-vector-meson dominance due to the axial anomaly.
\end{abstract}

\vfill

\end{titlepage}
\setcounter{footnote}{0}

\baselineskip 18pt \pagebreak
\renewcommand{\thepage}{\arabic{page}}
\pagebreak

\section{Introduction}

The idea of holography in studying difficult strong coupling dynamics of large $N_c$ gauge theories
has given us many useful insights and sometimes practical results that would have been much harder to obtain by conventional field theoretical
techniques. One example is large $N_c$ QCD whose importance cannot be emphasized too much.
There have appeared several approaches to construct holographic dual model of QCD \cite{Sakai:2004cn,Erlich:2005qh,Da Rold:2005zs,Casero:2005se,Gursoy:2007cb}, coined as holographic QCD, and 
it is an important open problem to improve the current models. 

Another avenue one can explore is to analyze lower dimensional QCD for which we may have a better control on the quantum field theory
side \cite{Hong:2010sb}. The motivation is manifold: 1) Test holographic results against known field theoretical results say in two dimensional QCD; 2) Seek 
new insights from holography; 3) Understand the role of matter such as temperature and density on lower dimensional hadronic spectra at large
$N_c$; 4) Extend our understanding to holographic condensed matter models.

In this paper, we mainly study two dimensional large $N_c$ massless QCD, or the t'Hooft model with massless quarks 
in the framework of holography. We are aware of two previous studies ~\cite{Gao:2006up,Rodriguez:2005jr}  whereby this
problem was addressed with incorrect conclusions. Indeed, in both analyses the key role played by the two-point axial
anomaly on the dynamics of massless fermions through the three dimensional Chern-Simons term was overlooked.  While
for four dimensional QCD the Chern-Simons term is subleading compared to the leading Maxwell term in the probe brane action, 
in two-dimensions it is at par with the Maxwell contribution, thereby altering the holographic renormalization and the meson spectra. 
A similar observation with different motivations appeared recently in~\cite{Kraus:2006wn,Davis:2007ka,
Balasubramanian:2010sc,Jensen:2010em}.

We will also address the issue of mass generation of photons in two dimensions or the Schwinger-mechanism
by coupling a dynamical vector-like gauge potential to the flavor vector current. The holographic  analysis involves only the 
three dimensional Chern-Simons term irrespective of the details of the massive mesonic spectrum and is therefore quite robust. 
As a result, the photon in two dimensions mingles only with the massless pseudoscalar in the massless quark limit thanks to 
the anomaly. This anti-vector-meson dominance behaviour is expected to be relaxed in the massive case.

The model we study is based on  a top-down construction using D-brane embeddings which parallels the construction by Witten~\cite{Witten:1998zw} and Sakai-Sugimoto for four dimensional QCD~\cite{Sakai:2004cn}, and is first introduced  in~\cite{Gao:2006up}.
(See also~\cite{Katz:2007br} for a bottom-up, synthetic approach.) We consider $N_c$ $D2$ branes compactified on a circle with anti-periodic boundary condition for fermions to realize two dimensional Yang-Mills theory at low energy. Fundamental massless fermions are introduced by inserting probe $D8$ and $\overline{D8}$ branes, in analogy with domain wall fermions.  
One drawback of the $D2$ background is that it contains dynamical glueballs in the spectrum which are  absent
in two dimensional Yang-Mills theory \cite{ictp}. However,  we expect this concern to be minor on the $D8$ brane flavor
dynamics.  This expectation is supported by our results.

In section 2 we give explicit construction of the holographic model for massless two-dimensional QCD at large $N_c$.  The emergence of the
Chern-Simons term which is at par with the Maxwell term is emphasized. In section 3,  we detail the near boundary asymptotics
for the holographic gauge field and work out the details of the holographic renormalization. In section 4, we analyze the flavored meson 
spectrum, which is shown to transmute to the chiral t'Hooft spectrum for a pertinent choice of the compactification circle. 
In section 5, we discuss how our results fit with the field-theoretical expectations at large $N_c$ with massless quarks.
In section 6,  we show how the Schwinger mechanism emerges from our analysis when external but dynamical photons are coupled to the flavor vector current. Our conclusions are in section 7.

\section{Holographic two dimensional QCD: $D2/D8/\overline{D8}$}

In this section, we summarize the set-up of the model introduced in~\cite{Gao:2006up}.
The construction is well-known and parallels that of the higher dimensional Sakai-Sugimoto model \cite{Sakai:2004cn}. To realize
1+1 dimensional Yang-Mills theory, one compactifies $N_c$ D2 branes on $S^1$ with anti-periodic boundary condition for fermions,
so that below the compactification scale $M_{KK}$, the only massless degrees of freedom are Yang-Mills gauge fields \cite{Witten:1998zw}.
The corresponding gravity solution is easily obtained by double Wick rotation from a non-extremal solution.
A useful reference for the solution with a consistent convention is~\cite{Itzhaki:1998dd}, from which one has
\bear
ds^2_{10D}&=& \left(r\over R\right)^{5\over 2}\left(\eta_{\mu\nu}dx^\mu dx^\nu +f(r)\left(dx^2\right)^2\right)
+\left(R\over r\right)^{5\over 2}\left({dr^2\over f(r)}+r^2 d\Omega_6^2\right)\quad,\nonumber\\
f(r)&=& 1-\left(r_{KK}\over r\right)^5\quad,\quad r_{KK}^{3\over 2}={2\over 5} M_{KK} R^{5\over 2}\quad,\quad
R^5=6\pi^2g_s N_c l_s^5\quad,\nonumber\\
e^\phi&=& g_s \left(R\over r\right)^{5\over 4}\quad,\quad {1\over \left(2\pi l_s\right)^5}\int_{S^6}\, F^{RR}_6 = N_c\quad,
\label{metric}
\eear
where the $D2$ branes are compactified along $x^2\sim x^2+{2\pi\over M_{KK}}$, and the 1+1 dimensional QCD coupling constant 
at the scale of $M_{KK}$ is given by
\bear
g^2_{YM}={g_s M_{KK}\over 2\pi l_s}\quad.
\eear
As pointed out in the introduction, this background contains dynamical colorless excitations, at variance
with 1+1 dimensional Yang-Mills theory with no dynamical spectrum~\cite{ictp}.  These unwanted glueballs
are likely the relics of the adjoint massive fields above $M_{KK}$. They will not interest us and therefore will
be ignored throughout. 

We will focus on the flavored meson sector by introducing probe $D8/\overline{D8}$ branes in the
background (\ref{metric}) such that they intersect $D2$ branes in 1+1 dimensions, assuming that the flavor sector remains immune to the adjoint but massive scalars
at low energy. See \cite{Jokela:2011eb} for a different possible embedding relevant to 1+2 dimensional theory. This assumption is borne out by the relative success of the Sakai-Sugimoto model in higher 
dimensions. We also note that unlike the $D4/\overline{D4}$ and $D6/\overline{D6}$, the $D8/\overline{D8}$
supports flux-vortex-instantons in the bulk for the $S^6$-wrapped $D6$ branes much like in the Sakai-Sugimoto model. These are the holographic
precursors of baryons at large $N_c$.

We will consider the maximally separated $D8/\overline{D8}$ branes embedding, thereby penalizing 
the exchange of massive adjoint fields above $M_{KK}$. Because of the bulk gravity geometry,
the probe $D8/\overline{D8}$ are fused at the tip of $r=r_{KK}$ (cigar-like).  This configuration geometrically breaks
chiral symmetry as it ties the left($D8$)/right($\overline{D8}$)-handed probe branes. While the spontaneous breaking in
1+1 dimensions is ruled out by the Mermin-Wegner-Coleman theorem, it is not at large $N_c$ where
a would-be Goldstone boson emerges~\cite{ARIEL}. The finite $1/N_c$ corrections are known to 
wash out the long range order in 1+1 dimensions, causing all correlation functions to fall-off as power laws at
large distances. This long range order-disorder transition at $N_c=\infty$ can be mapped to the BKT-type 
(Berezinskii-Kosterlitz-Thouless) transition. However, it is outside the scope of the leading $N_c$
holographic approach we are pursuing.


With these in mind, the probe $D8$-brane action is given by
\bear
S_{D8}=-\mu_8 \int d^9\xi\, e^{-\phi} \sqrt{{\rm det}\left(g^*+2\pi l_s^2 F\right)}
+\mu_8 {\left(2\pi l_s^2\right)^2 \over 2!}\int\, F^{RR}_6\wedge A\wedge F\quad,
\eear
where $\mu_p=(2\pi)^{-p} l_s^{-(p+1)}$ is the tension of $p$-brane. For $\overline{D8}$ brane action one flips the sign of the Chern-Simons term. 
Our $D8$ brane wraps $(x^0,x^1,r,S^6)$, and a
straightforward but tedious computation gives us the quadratic action of the world-volume gauge field as
\bear
S^{(2)}_{D8}&=&{N_c\over 10\pi}\int d^2x dr\, \left(r\over f(r)^{1\over 2}\right)
{1\over 2}\left(f(r) F_{\mu r}F^\mu_{\,\,\,\,r} -{1\over 2}\left(R\over r\right)^5 F_{\mu\nu}F^{\mu\nu}\right)\nonumber\\
&+&{N_c\over 8\pi}\int d^2x dr \, \epsilon^{MNP}A_M F_{NP}\quad,
\eear
where $\mu,\nu=0,1$. 
Each $D8$ and $\overline{D8}$ brane spans radial direction $r_{KK}\le r <\infty$, and they join at the tip $r=r_{KK}$.
Following the literature, let's introduce a dimensionless variable $z$ defined by
\be
\left( r\over r_{KK}\right)^5=1+z^2\quad,
\ee
and assign $z>0$ for $D8$ and $z<0$ for $\overline{D8}$ so that $-\infty< z <\infty$ spans the two branes simultaneously.
Upon this change of variable, the total action for both $D8$ and $\overline{D8}$ branes reads as
\bear
S^{(2)}_{D8/\overline{D8}}&=&
{N_c\over 8\pi}\int d^2x dz\, \left(\left(1+z^2\right)^{1\over 2}F_{\mu z}F^\mu_{\,\,\,\,z} 
-{1\over M_{KK}^2}\left(1+z^2\right)^{-{11\over 10}}{1\over 2} F_{\mu\nu}F^{\mu\nu} \right)\nonumber\\
&+&{N_c\over 8\pi}\int d^2 x dz \, \epsilon^{MNP} A_M F_{NP}\quad,\label{start}
\eear
with a single sign of the Chern-Simons term valid for all $z$.

Our convention is $\epsilon^{\mu\nu z}=\epsilon^{\mu\nu}$, $\epsilon^{01}=+1$, and $\eta_{\mu\nu}={\rm diag}(+1,-1)$.
Note that upon introducing $z$, we should also change the radial component of gauge field as $A_r dr=A_z dz$.
The above action is our starting point for the study of the  flavored meson spectrum, paying special attention to
the existence and proper identification of massless pseudo scalar state which is mandated by chiral symmetry.
For simplicity, we consider a single flavor $N_F=1$ as the generalization to the case of multi-flavors is straightforward.

Our main emphasis will be on the importance of the 3D Chern-Simons term in the analysis of the meson spectrum of the theory, contrary
to higher dimensional examples in previous literatures.
There are several reasons for this. First, the 3D Chern-Simons term is quadratic in the gauge fields just like the usual Maxwell term, so
that it affects the 2-point propagators that determine the spectrum. In fact, it has been known for a while
that 3D Maxwell-Chern-Simons theory describes a massive spin 1 particle. Being massive, its near boundary asymptotics
around the UV region is completely different from the massless pure Maxwell theory. Indeed, we will see in the next section that this
yields different holographic renormalization and boundary counter terms.

The previous consideration in~\cite{Rodriguez:2005jr} has not taken into account the 
3D Chern-Simons term, and as a result their analysis found a serious difficulty in computing the meson spectrum,
i.e. the absence of a normalizable massless state as expected by the spontaneous breaking of chiral symmetry
due to a strange logarithmic divergence. Also the choice of the type of boundary condition for the massive states was
not clear. We will see that these problems are mere artifacts of not including the 3D Chern-Simons term in the spectrum 
analysis. Being massive with the 3D Chern-Simons term, the correct near boundary
power-law behavior without any logarithms makes the boundary condition clearer.

Another important point is the chirality at both boundaries $z=\pm \infty$. 
According to the holographic dictionary, one would expect the gauge field at one boundary, say $A^\mu(z\to+\infty)$, to
couple to the left-handed field theory current, $j_\mu^{(L)}$, while at the other boundary $A^\mu(z\to-\infty)$ to $j_\mu^{(R)}$.
Specific to 1+1 dimensions, the components of these chiral currents are not all independent due to a special property of  the
1+1 dimensional gamma matrices. They are constrained algebraically as
\bear
j_\mu^{(L)}=- \epsilon_{\mu\nu}j^{\nu(L)}\quad,\quad j_\mu^{(R)}=+\epsilon_{\mu\nu}j^{\nu(R)}\quad,
\label{2D}
\eear
or more explicitly $j^{(L)}_0=-j^{(L)}_1$, $j^{(R)}_0=+j^{(R)}_1$. This algebraic constraint implies that
the sources $A^\mu(\pm\infty)$ for the chiral currents should also be constrained similarly, which is not obvious a priori.
What happens with the Chern-Simons term is that the two chiral components of $A^\mu(+\infty)$ behave differently near $z\to\infty$,
so that only one component is selected as a source for the boundary current, while the other component will be interpreted differently
as we will expound in the next section. The same things happen near the other boundary $z\to -\infty$ with the chirality flipped.
The 3D Chern-Simons term therefore plays an essential role in making the dual theory consistent with the chirality of the  currents.

Finally, looking at (\ref{start}) one notices that the Maxwell and Chern-Simons terms are both leading in $N_c$ without any 
relative suppression of $\lambda=g_{YM}^2 N_c$ or $N_c$. This is an interesting difference from the Sakai-Sugimoto model where
one finds that 5D Chern-Simons terms are $1\over \lambda$-supressed compared to the Maxwell term. 
This tells us that the 3D Chern-Simons term in a holographic description of 2D QCD
is an important leading order ingredient that has to be included in the spectrum analysis. In 1+1 dimensions
the chiral anomaly affects the spectrum in a fundamental way.


\section{Holographic renormalization \label{sec3}}

In this section we present the holographic renormalization of the theory, which will be important for correctly identifying 
the pseudo scalar states in the later sections.
One starts with the equations of motion,
\bear
\partial_z\left(\left(1+z^2\right)^{1\over 2} F^{\mu}_{\,\,\,\,z}\right)-{1\over M_{KK}^2}\left(1+z^2\right)^{-{11\over 10}}
\partial_\nu F^{\mu\nu} +2 \epsilon^{\mu\nu} F_{\nu z} &=& 0\quad,\nonumber\\
\left(1+z^2\right)^{1\over 2} \partial_\mu F^{\mu}_{\,\,\,\,z} - \epsilon^{\mu\nu} F_{\mu\nu}&=& 0\quad.
\eear
The chirality projection operators are
\be
P^{(\pm)}_{\mu\nu}={1\over 2}\left(\eta_{\mu\nu}\pm \epsilon_{\mu\nu}\right)\quad,
\ee
which obey
\be
\epsilon^{\mu\nu}P^{(\pm)}_{\nu\alpha}=\pm P^{(\pm)\mu}_{\quad\,\,\,\,\,\,\alpha}\quad,\quad
P^{(\pm)}_{\mu\nu}=P^{(\mp)}_{\nu\mu}\quad,\quad P^{(\pm)}_{\mu\nu}P^{(\pm)\nu\alpha}=P^{(\pm)\alpha}_\mu\quad,\quad
P^{(\pm)}_{\mu\nu}P^{(\mp)\nu\alpha}=0\quad,
\ee
and the completeness relation $\eta_{\mu\nu}=P^{(+)}_{\mu\nu}+P^{(-)}_{\mu\nu}$.
We will work in the axial gauge $A_z=0$. Due to the completeness of the chiral projection operators,
one can generally decompose $A_\mu$ as $A_\mu=P^{(+)}_{\mu\nu}A^\nu+P^{(-)}_{\mu\nu}A^\nu\equiv f^{(+)}_\mu + f^{(-)}_\mu$,
in terms of which the equations of motion become
\bear
\partial_z\left(\left(1+z^2\right)^{1\over 2}\partial_z f^{(+)}_\mu \right)
+{1\over M_{KK}^2} \left(1+z^2\right)^{-{11\over 10}}\left( -{1\over 2} \partial_\nu\partial^\nu f^{(+)}_\mu +
\partial_\mu^{(+)}\partial_\nu^{(+)}f^{(-)\nu}\right)+2 \partial_z f^{(+)}_\mu &=& 0,\nonumber\\
\partial_z\left(\left(1+z^2\right)^{1\over 2}\partial_z f^{(-)}_\mu \right)
+{1\over M_{KK}^2} \left(1+z^2\right)^{-{11\over 10}}\left( -{1\over 2} \partial_\nu\partial^\nu f^{(-)}_\mu +
\partial_\mu^{(-)}\partial_\nu^{(-)}f^{(+)\nu}\right)-2 \partial_z f^{(-)}_\mu &=& 0,\nonumber\\
\left(\left(1+z^2\right)^{1\over 2}\partial_z+2\right)\partial^{(-)}_\mu f^{(+)\mu} +
\left(\left(1+z^2\right)^{1\over 2}\partial_z-2\right)\partial^{(+)}_\mu f^{(-)\mu}&=& 0,\nonumber\\ 
\eear
with the chiral derivative $\partial^{(\pm)}_\mu\equiv P^{(\pm)}_{\mu\nu}\partial^\nu$. We used the identities
\be
\partial^{(\pm)}_\mu \partial^{(\mp)}_\nu f^{(\pm)\nu}={1\over 2}\partial_\nu\partial^\nu f^{(\pm)}_\mu\quad.
\ee

To discuss the near boundary asymptotics, it is convenient to introduce $w=\sinh^{-1}(z)$ so that $\left(1+z^2\right)^{{1\over 2}}\partial_z
=\partial_w$, in terms which the equations of motion are
\bear
\partial_w^2 f^{(+)}_\mu 
+{1\over M_{KK}^2} \left(\cosh w\right)^{-{6\over 5}}\left( -{1\over 2} \partial_\nu\partial^\nu f^{(+)}_\mu +
\partial_\mu^{(+)}\partial_\nu^{(+)}f^{(-)\nu}\right)+2 \partial_w f^{(+)}_\mu &=& 0,\nonumber\\
\partial_w^2 f^{(-)}_\mu 
+{1\over M_{KK}^2} \left(\cosh w\right)^{-{6\over 5}}\left( -{1\over 2} \partial_\nu\partial^\nu f^{(-)}_\mu +
\partial_\mu^{(-)}\partial_\nu^{(-)}f^{(+)\nu}\right)-2 \partial_w f^{(-)}_\mu &=& 0,\nonumber\\
\left(\partial_w+2\right)\partial^{(-)}_\mu f^{(+)\mu} +
\left(\partial_w-2\right)\partial^{(+)}_\mu f^{(-)\mu}&=& 0,\label{weq}
\eear
from which one can study the near boundary $w\to+\infty$ asymptotics and the necessary
counterterms to regulate the emerging divergences. The opposite boundary, $w\to-\infty$, has 
an identical structure with $w\leftrightarrow -w$
and $(+)\leftrightarrow (-)$. After some algebra, we have the following expansion near $w\to+\infty$,
\bear
f_\mu^{(+)}&=& A_\mu^{(+)}+c_\mu^{(+)(1)}e^{{4\over 5} w}+c_\mu^{(+)(2)}e^{-{2\over 5} w}+c_\mu^{(+)(3)}e^{-{6\over 5} w}+
c_\mu^{(+)(4)}e^{-{8\over 5} w}+\tilde c_\mu^{(+)}e^{-2w}+\cdots\quad,\nonumber\\
f_\mu^{(-)}&=& A_\mu^{(-)}+c_\mu^{(-)(0)}e^{2w}+ c_\mu^{(-)(1)}e^{{4\over 5} w}+c_\mu^{(-)(2)}e^{-{2\over 5} w}+
\cdots+\tilde c_\mu^{(-)}e^{-{16\over 5}w}+\cdots\quad,\label{exp}
\eear
where the leading $c^{(-)(0)}_\mu$ term is unconstrained. This should be interpreted as a source 
for some boundary chiral operator ${\cal O}^{(+)}_\mu$. Note that ${\cal O}^{(+)}_\mu$ is not expected to be the chiral current
operator $j^{(+)}_\mu$ because it has a nontrivial dimension due to $e^{2w}$ factor. We will elaborate further on ${\cal O}^{(+)}_\mu$ later.
The subleading terms are completely fixed by $c^{(-)(0)}_\mu$ until it gets to
$\tilde c_\mu^{(+)}e^{-2w}$ which is also unconstrained and corresponds to a normalizable mode. This normalizable mode also appears in
the $(-)$ chirality starting from $\tilde c_\mu^{(-)}e^{-{16\over 5}w}$.
Explicitly,
\bear
c_\mu^{(+)(1)}&=&-{25\over 56}\left(2^{6\over 5}\over M_{KK}^2\right) \partial_\mu^{(+)}\partial_\nu^{(+)} c^{(-)(0)\nu}\quad,
\quad c_\mu^{(-)(1)}=-{25\over 48}\left(2^{6\over 5}\over M_{KK}^2\right) \Box c^{(-)(0)}_\mu\quad,
\nonumber\\
c_\mu^{(+)(2)}&=&-{625\over 1344}\left(2^{6\over 5}\over M_{KK}^2\right)^2 \Box 
\partial_\mu^{(+)}\partial_\nu^{(+)} c^{(-)(0)\nu}\quad,\quad 
c_\mu^{(-)(2)}=-{625\over 4032}\left(2^{6\over 5}\over M_{KK}^2\right)^2 \Box^2 c^{(-)(0)}_\mu\quad,\nonumber\\
c_\mu^{(+)(3)}&=&-{5\over 4}\left(2^{6\over 5}\over M_{KK}^2\right) \partial_\mu^{(+)}\partial_\nu^{(+)} c^{(-)(0)\nu}\quad,
\quad c_\mu^{(+)(4)}= {15625\over 129024}\left(2^{6\over 5}\over M_{KK}^2\right)^3 \Box^2 \partial_\mu^{(+)}\partial_\nu^{(+)} c^{(-)(0)\nu}\quad,
\nonumber\\
\eear
where $\Box\equiv \partial_\mu \partial^\mu$.
What is unusual in the above is the constant mode $A_\mu^{(\pm)}$ which in fact has to be a pure gauge due to the last equation 
of (\ref{weq}); $\epsilon^{\mu\nu} F_{\mu\nu}=\partial_\mu^{(-)} A^{(+)\mu}-\partial_\mu^{(+)} A^{(-)\mu}=0$.
Normally one would gauge them away but in the presence of the 3D Chern-Simons term, they become dynamical due to the non-gauge invariance of the 
Chern-Simons term, and for the case of pure Chern-Simons theory they indeed give dynamical degrees of freedom of WZW theory on the boundary \cite{Jensen:2010em,Elitzur:1989nr}. Based on this, one expects to find a massless scalar in this sector which is a bosonization of the chiral 
currents, to which we come back in section \ref{schwinger}.

The bulk action in our gauge is 
\bear
S_{\rm bulk}&=& {N_c\over 8\pi}\int d^2 x dw\,\left(\left(\partial_w A_\mu\right)\left(\partial_w A^\mu\right)-{1\over M_{KK}^2}
\left(\cosh w\right)^{-{6\over 5}} {1\over 2}F_{\mu\nu} F^{\mu\nu} -2 \epsilon^{\mu\nu}A_\mu \left(\partial_w A_\nu\right)\right),\nonumber
\label{newaction}
\eear
and after a long computation with the expansion (\ref{exp}) one obtains divergences near $w\to+\infty$ as (we omit the common integral symbol $\int d^2x$ in the following)
\bear
\left(8\pi\over N_c\right)S_{\rm div}&=&{5\over 4}\left(2^{6\over 5}\over M_{KK}^2\right)\left(\partial_\mu^{(+)} c^{(-)(0)\mu}\right)^2 e^{{14\over 5}w}
+2 A_\mu^{(+)} c^{(-)(0)\mu}e^{2w} \nonumber\\
&+&{125\over336}\left(2^{6\over 5}\over M_{KK}^2\right)^2\left(\partial_\mu^{(+)} c^{(-)(0)\mu}\right)\Box
\left(\partial_\nu^{(+)} c^{(-)(0)\nu}\right) e^{{8\over 5}w}\nonumber\\
&+&  \left(2^{6\over 5}\over M_{KK}^2\right)\left(\left(\partial_\mu^{(+)} c^{(-)(0)\mu}\right)^2-{25\over 42}
A_\mu^{(+)} \Box c^{(-)(0)\mu}\right) e^{{4\over 5}w} \nonumber\\
&-&{78125\over451584}\left(2^{6\over 5}\over M_{KK}^2\right)^3 \left(\Box \partial_\mu^{(+)} c^{(-)(0)\mu}\right)^2 e^{{2\over 5}w}\quad,\label{div}
\eear
which needs to be cancelled by introducing boundary counterterms. 
The Maxwell term in the original bulk action in the above can be written as 
\bear
-\int d^3x \,\sqrt{-g}{1\over 4e^2}F_{MN}F_{PQ}g^{MP}g^{NQ}\quad,
\eear
with the effective metric $g_{MN}dx^M dx^N=g(w)\left(\left(\cosh w\right)^{6\over 5}\eta_{\mu\nu}dx^\mu dx^\nu +M_{KK}^{-2}dw^2\right)$ and
the coupling constant $e^{-2}={N_c\over 4\pi M_{KK}}\sqrt{g(w)}$ with an arbitrary function $g(w)$,
so that one can take the induced 1+1 dimensional metric on the boundary as
\bear
\gamma_{\mu\nu}=\left(\cosh w\right)^{6\over 5}g(w)\eta_{\mu\nu}\quad.
\eear
Regarding the freedom of $g(w)$, if we want $e^{-2}$ to become a constant at large $w$, we require $g(w)\to 1$ as $w\to\infty$.
We will mention more about $g(w)$ shortly.
The possible covariant counterterms one can consider are
\bear
\left(8\pi\over N_c\right) S_{\rm counter}&=& \sqrt{-\gamma}\Bigg(C_1 A_\mu A_\nu \gamma^{\mu\nu}+
{C_2\over M_{KK}^2} A_\mu \Box_\gamma A_\nu \gamma^{\mu\nu}+{C_3\over M_{KK}^4} A_\mu \Box_\gamma^2 A_\nu \gamma^{\mu\nu}\label{counters}\\ 
&+&
{C_4\over M_{KK}^2} F_{\alpha\beta}F_{\delta\eta}\gamma^{\alpha\delta}\gamma^{\beta\eta}
+{C_5\over M_{KK}^4} F_{\alpha\beta}\Box_\gamma F_{\delta\eta}\gamma^{\alpha\delta}\gamma^{\beta\eta}
+{C_6\over M_{KK}^6} F_{\alpha\beta}\Box_\gamma^2 F_{\delta\eta}\gamma^{\alpha\delta}\gamma^{\beta\eta}\Bigg),\nonumber
\eear
where $\Box_\gamma=\partial_\mu\partial_\nu \gamma^{\mu\nu}$ and higher derivative terms can easily be shown to be irrelevant for divergences.
As the independent components in (\ref{div}) are six, and we seem to have six free parameters in (\ref{counters}), one might expect a unique solution for $C_i$, but explicit computations show that the three types of divergences with $e^{{14\over 5}w},e^{2w}$ and the first of $e^{{4\over 5}w}$ are from $C_1,C_4$. The problem of fixing the $C_i$'s is over-constrained. Requiring that there exists a solution fixes our freedom for $g(w)$ uniquely such that
\be
g(w)=1+3 e^{-2|w|}+\cdots\quad.
\ee
Also, $C_3,C_6$ contribute to only the $e^{{2\over 5}w}$ divergence, so only a linear combination of $C_3$ and $C_6$
is determined. The result after some algebra is
\be
C_1=-1\quad,\quad C_2=-{25\over56}\quad,\quad C_4={5\over 28}\quad,\quad C_5=-{125\over 896}\quad,\quad
C_3 -{56\over 25}C_6=-{8375\over 18816}\quad.
\ee

It is worth mentioning that the first three terms in (\ref{counters}) appear to be gauge non-invariant, which is a common feature of
holography with a 3D Chern-Simons term. Indeed, the term

\be
-{N_c\over 8\pi} \int d^2x\, \sqrt{\gamma}  A_\mu A_\nu \gamma^{\mu\nu} \Bigg\vert_{z={\pm{1\over \epsilon}}}\quad,
\label{1counter}
\ee
already appeared in the pure Chern-Simons theory in~\cite{Elitzur:1989nr}, and its relevance to holography
was pointed out in~\cite{Davis:2007ka,Jensen:2010em}. Note that this term is independent of $g(w)$, and fixed only
by the $e^{2w}$ divergence. We recall that the pure Maxwell theory has a completely different counterterm \cite{Hung:2009qk,Ren:2010ha}.

The Maxwell-Chern-Simons theory is closer to the pure Chern-Simons theory rather than the pure Maxwell theory. One way to see
this is to continuously vary the coefficient of the Maxwell term to vanish, as the coefficient of Chern-Simons term is fixed to its quantized
integer value. Indeed, if we set $M_{KK}\to\infty$ (low-energy limit) we recover the boundary term of the pure Chern-Simons theory. 
Note that the same boundary term arises from the structure of the divergences, which is not a priori related to the pure Chern-Simons theory consideration as for example in~\cite{Jensen:2010em}.
The next step is to compute the holographic expectation value of the operator that the bulk gauge field couples to by
varying the total regularized action $S_{\rm bulk}+S_{\rm counter}$ with respect to the boundary value of $A_\mu$.

For the $(-)$ chirality, it is clear that the leading $c_\mu^{(-)(0)}$ is a source to some boundary operator ${\cal O}_\mu^{(+)}$ at $w\to+\infty$ that is dual to the massive sector of the bulk theory. The normalizable mode, $\tilde c^{(+)}_\mu$, is naturally the expectation value of ${\cal O}_\mu^{(+)}$.
(A similar identification works for the opposite boundary $w\to-\infty$.) Its dimension is encoded in its wavefunction $e^{2w}$ which is unusual 
for massless chiral currents in 1+1 dimensions. Therefore,  ${\cal O}_\mu^{(+)}$ should be some other chiral composite operator of the theory. 
In the next section, we show that massive meson spectrum arises from the $w$-dependent modes suggesting that the operators ${\cal O}_\mu^{(\pm)}$ are
associated to these states.

For the $(+)$ chirality near the $w\to\infty$ boundary expansion, the constant mode $A_\mu^{(+)}$ is another
leading unconstrained mode that is independent of $c_\mu^{(-)(0)}$. It should be treated as a separate source to some other boundary operator.
Since the wave function is constant (which is similar to the higher dimensional example) it is natural to associate it with the boundary chiral current $j^{(-)}_\mu$ that lives on the $w\to\infty$ boundary. The interesting point is how the bulk Chern-Simons term selects the chirality of the source, which is needed because the boundary current operator is also chiral.  The other constant mode  $A_\mu^{(-)}$, is the expectation value of $j^{(-)}_\mu$. The same conclusion for the pure Chern-Simons theory was obtained in~\cite{Jensen:2010em,Elitzur:1989nr}.  It is interesting that we get to the same conclusion by including the massive modes and considering divergences. The reverse holds at the opposite boundary $w\to -\infty$. 
A detailed computation from our $S_{\rm bulk}$ and $S_{\rm counter}$ yields at $w\to +\infty$,
\be
\langle {\cal O}^{(+)}_\mu \rangle= -{N_c \over 2\pi} \,\,\tilde c_\mu^{(+)}\quad,\quad
\langle j^{(-)}_\mu\rangle = -{N_c\over 2\pi} \,\,A^{(-)}_\mu\quad,\label{expec}
\ee
and similarly for $w\to-\infty$.

The fact that two independent bulk/boundary correspondences are retained for a single gauge field in the bulk, maybe understood
from the fact that the 3D Maxwell-Chern-Simons theory contains two sectors: a massive spin 1 sector and a massless pure gauge,
both of which are kept in our case. A same observation was made in~\cite{Gukov:2004id}.
A related fact that we will come back to in the next section is that the theory contains a completely decoupled sector; the mode  $A^{(\pm)}_\mu$ which is contant in the entire $w$ range and is a pure gauge.  This mode is expected to contain the massless scalar one expects from the boundary chiral dynamics of the WZW model and is governed only by the 3D Chern-Simons term.
What is puzzling in our situation of having two boundaries is that the source/expectation value dictionary is reversed in the two boundaries $w\to \pm\infty$, so that $A^{(\pm)}_\mu$ are naively both sources and expectation values. Typically one would turn off sources to study dynamical modes
of the theory, but this would kill the constant mode completely, which shouldn't be the case. We will resolve this puzzle in section {\ref{schwinger}},
and the upshot will be that we can keep it as a dynamical mode of the theory.

\section{Flavored meson spectrum \label{sec4}}  

We now study the meson spectrum of the theory, including the 3D Chern-Simons term as an essential ingredient of the model. Since the
boundary terms upset gauge-invariance, we choose to work with equations of motion which are gauge-invariant. We work in the axial
gauge $A_w=0$.
Working with the equations of motion in the spectrum analysis means that one assumes a definite energy-momentum mode $e^{-i p\cdot x}$ with
$p^2=m_{(n)}^2$, where $m_{(n)}^2$ is the mass square one would like to determine.
By imposing the normalizable boundary condition at $w\to\pm\infty$, the mode spectrum with $p^2=m_{(n)}^2$ is discrete.

Our starting point is (\ref{weq}) assuming a common $e^{-ip\cdot x}$ factor in the wave functions.
Inspecting the Lorentz index structure, one arrives uniquely at the following Ansatz for the bulk wave function profile,
\be
A_\mu=f^{(+)}_\mu+f^{(-)}_\mu = p^{(+)}_\mu f^{(+)}(w)+p^{(-)}_\mu f^{(-)}(w)\quad,\label{ansatz}
\ee
where $p^{(\pm)}_\mu = P_{\mu\nu}^{(\pm)}p^\nu$ are the chirally projected momenta. 
The decomposition (\ref{ansatz}) is different from the one encountered in higher dimensions Sakai-Sugimoto~\cite{Sakai:2004cn},
and is due to our inclusion of the 3D Chern-Simons in the equations of motion.
Inserting the above into (\ref{weq}) and using $p^2=2p_\mu^{(+)} p^{(-)\mu}$ etc, one arrives at
\bear
\partial_w^2 f^{(+)}+{1\over M_{KK}^2}\left(\cosh w\right)^{-{6\over 5}} {1\over 2} p^2 \left(f^{(+)}-f^{(-)}\right)
+2\partial_w f^{(+)}&=&0\quad,\nonumber\\
\partial_w^2 f^{(-)}+{1\over M_{KK}^2}\left(\cosh w\right)^{-{6\over 5}} {1\over 2} p^2 \left(f^{(-)}-f^{(+)}\right)
-2\partial_w f^{(-)}&=&0\quad,\nonumber\\
\left(\partial_w+2\right)f^{(+)} +\left(\partial_w -2\right) f^{(-)} &=& 0\quad,
\eear
where adding the first two equations is trivially satisfied by the third equation, so that there are only two independent equations.
Subtracting the first two equations and using the third equation which can be recast as
\be
\partial_w\left(f^{(+)}+f^{(-)}\right)=-2\left(f^{(+)}-f^{(-)}\right)\quad,\label{third}
\ee
one obtains a single second order differential equation for $\left(f^{(+)}-f^{(-)}\right)$  only, which is
\be
\left(\partial_w^2 +{1\over M_{KK}^2}\left(\cosh w\right)^{-{6\over 5}}p^2 -4 \right) \left(f^{(+)}-f^{(-)}\right)=0\quad.\label{eigen}
\label{EQ}
\ee
The natural  boundary condition at $w\to\pm\infty$ is $e^{\mp 2w}$ for the normalizable modes as identified earlier.
This is a well-defined eigenvalue problem for a discrete mass square $p^2=m_{(n)}^2$.
It is useful to note that the combination $\left(f^{(+)}-f^{(-)}\right)$ in (\ref{ansatz}) corresponds to $\epsilon^{\mu\nu}F_{\mu\nu}$ and
hence is gauge invariant.
As the equation (\ref{eigen}) is even under $w\to-w$, the eigen functions are either even or odd under $w\to -w$.

Our numerical determination of the spectrum is
\be
{m^2\over M_{KK}^2}=5.515(+),\quad 9.623(-),\quad 14.687(+),\quad 20.705(-), \quad\cdots
\label{NUM}
\ee
where $(\pm)$ denotes even or odd under $w\to-w$. Parity $P$ in two dimensions maps to $w\leftrightarrow -w$
and $f^{(+)}\leftrightarrow f^{(-)}$ in our holographic model, hence $P=-(\pm)$. The lowest mode is a vector and the next is a pseudovector, 
and so on. Charge conjugation $C$ corresponds to $w\to-w$ and $A_\mu\to -A_\mu^T=-A_\mu$~\cite{Sakai:2004cn}, and one finds that it is not a symmetry of the 3D Chern-Simons term. In fact, the 2-point axial anomaly in two dimensions $\partial_\mu j^\mu_A \sim \epsilon_{\mu\nu}F_{V}^{\mu\nu}$ violates $C$ maximally. Hence $C$ is not a good symmetry to talk about.
One can understand this from the view-point of dimensional reduction of 4D to 2D in a very large magnetic field \cite{Basar:2010zd,Kharzeev:2010gd} as the magnetic field is $P$-even and $C$-odd. 

Once $\left(f^{(+)}-f^{(-)}\right)$ is found, one integrates (\ref{third}) to find $\left(f^{(+)}+f^{(-)}\right)$, which completes the solution.
One subtlety is the integration constant that appears in integrating (\ref{third}), which corresponds to a constant shift of
$f^{(+)}$ and $f^{(-)}$ in the entire $w$ range by a same amount because $\left(f^{(+)}-f^{(-)}\right)$ is already fixed. Looking at our original Ansatz (\ref{ansatz}), this constant shift is a $w$-independent pure gauge, $A_\mu=\partial_\mu \Lambda$.
As we noted earlier, we would like to keep this pure gauge and constant mode to obtain the massless scalar state one expects
on the field theory side. This mode will be treated as a separate sector of the theory shortly.
In fact, one easily checks that $p^2=0$ cannot be solved as a normalizable solution, and
(\ref{eigen}) does not contain the sought-for massless state. In any case, our spectrum from (\ref{eigen}) is robust.

Another important point is to note that the difference in values of $\left(f^{(+)}+f^{(-)}\right)$ between $w\to\pm\infty$ is given by integrating (\ref{third}),  
\be
\Delta \left(f^{(+)}+f^{(-)}\right)= -2 \int_{-\infty}^{+\infty} dw'\,\left(f^{(+)}-f^{(-)}\right)(w')\quad,\label{mixing}
\ee
which will be non zero for even eigen functions of $\left(f^{(+)}-f^{(-)}\right)$.  
As $\left(f^{(+)}-f^{(-)}\right)$ vanishes at $w\to\pm\infty$, we have $f^{(+)}=f^{(-)}$ at $w\to\pm\infty$, and
the above is in fact a difference of the pure gauge contribution  at $w\to\pm\infty$,
\be
\partial_\mu \Lambda(x,w\to+\infty)- \partial_\mu \Lambda(x,w\to-\infty)\quad,
\ee
which is precisely similar to the pion field, or Wilson line along the $w$-direction, in higher dimensions~\cite{Sakai:2004cn}. 
Recall that in higher dimensional examples,
the pion appears in the $A_w=0$ gauge as the difference of the pure gauge terms between both boundaries $w\to\pm\infty$.
In our case, the above means that there is no separate pion field, but each massive vector of even wave function contains some component of it.
This is understood as follows: due to the 2-point axial anomaly in 1+1 QCD, the pion field has a mixed kinetic term with each naive
massive vector mesons so that the final diagonalized states are massive and contain some part of original pion in their wave functions.
Indeed, the $-2$ in the right-hand side of (\ref{mixing}) can be traced back precisely to the Chern-Simons coefficient.
The expected massless state has nothing to do with the original pion field which has become massive due to this mixing.
This is in line with the fact that even the free fermion theory transmutes to a massless scalar through bosonization.
This state has nothing to do with the spontaneous breaking of a symmetry.

To elaborate on this further, the decoupling of massless bosons from the massive vector mesons seems consistent with the picture derived
from the non-Abelian bosonization~\cite{Witten:1983ar}. The $N_c$ Dirac fermions with $N_F$ flavors are bosonized into
\be
N_c {\cal L}\left(U(N_F)\right) + N_F {\cal L}\left(SU(N_c)\right) \quad,
\ee
where ${\cal L}(G)$ is Wess-Zumino-Witten model of group $G$ with level 1. The two sectors of the theory above decouple 
from each other in the massless fermion limit. As one introduces $SU(N_c)$ Yang-Mills theory for the second sector, it provides various
massive vector mesons, while the first massless flavor sector remains intact and dominates at low energy.

\section{Comparison with 1+1 QCD}

The holographic approach we have pursued so far in the meson sector agrees qualitatively with the known, exact results
obtained from the field theoretical side either by resumming planar graphs or through non-Abelian bosonization.  It is worth
pointing out that our holographic approach is strictly for massless quarks $m=0$ and strong gauge coupling $\lambda=g^2 N_c$,
so it corresponds to the strong coupling limit $\lambda\gg m=0$. In this limit, the non-Abelian bosonization construction yields
 \cite{Witten:1983ar} that the chiral baryons are heavy~\cite{FRISHMAN}, and the anomaly is saturated by only the decoupled
chiral and massless mode. The same observation holds for the resummation of the planar graphs in the chiral limit~\cite{GROSS}
although a subtlety shows up since the large $N_c$, $\lambda$ limits and the chiral limit are not a priori inter-changeable. Also,
the 2-point vector-vector and axial-axial correlators are saturated by this chiral and massless boson in the massless limit,
a feature unique in 1+1 dimensions that can be traced back to the property (\ref{2D}). This is no longer true for finite but small quark mass $m$.

Finally, we observe that the numerical spectrum (\ref{NUM}) is approximately equi-distant with 
$m_{(n)}^2\approx n\,m_{(1)}^2$ or Regge-like.  For large $n$ a WKB analysis
of (\ref{EQ}) yields

\be
2\,\int^{(\alpha+\sqrt{\alpha^2-1})}_{1/(\alpha+\sqrt{\alpha^2-1})}
\,\frac{dx}x \left(\alpha\,\left(\frac{2x}{x^2+1}\right)^{6/5}-1\right)^{1/2}=(n+\frac 12)\pi
\label{WKB}
\ee
with $\alpha=m_{(n)}^2/4M^2_{KK}\geq 1$. While the WKB  threshold $m_{(n)}>2M_{KK}$ is consistent with the numerical analysis,
the WKB spectrum asymptotics for $n\gg 1$ is non-Regge-like with $m_{(n)}^2/M^2_{KK}\approx n^2/2$. It is tempting to suggest that the low
lying part of the mesonic spectrum is in agreement with the t'Hooft spectrum if we were to identify $M^2_{KK}\approx
\pi\lambda$. This identification is also supported by the alternance of the parity of the states, a signal that the fused $D8/\overline{D8}$ geometry mocks up the
spontaneous breaking of chiral symmetry.  The high lying and non-Regge-like mesonic states are then spurious as they are likely affected by the adjoint massive states. A more thorough comparison of the moments of the wavefunctions for the Regge-like states requires a finite quark mass
and will be detailed elsewhere.

\section{Chiral bosons and the Schwinger mechanism \label{schwinger}}

The salient features of the chiral massless boson arising from holography in 1+1 dimensions are: 
1)  Its emergence from the pure gauge sector of the bulk gauge fields; 2) Its complete decoupling
from the massive chiral states discussed earlier. While the former feature is similar to the 1+3 dimensional
holographic formulation~\cite{Sakai:2004cn}, the latter is not.  To understand the decoupling of the chiral
boson it is best to relax the axial gauge used in so far. Indeed, in general the pure bulk gauge modes are
defined through
\be
A_\mu=\partial_\mu \alpha(x,w)\quad,\quad A_w=\partial_w \alpha(x,w)\quad,
\ee
with an arbitrary function $\alpha(x,w)$. The presence of $A_w$ allows a non-trivial dependence of $A_\mu$ on $w$, 
which was absent in the $A_w=0$ gauge. From section {\ref{sec3}}, the near $w\to\infty$ value of 
$A_\mu^{(+)}(+\infty)=\partial_\mu^{(+)} \alpha(w=\infty)\equiv\partial_\mu^{(+)} \varphi_+$ should be identified as
the source for the boundary current $j^{(-)}_\mu$. From (\ref{expec})
\be
-{N_c\over 2\pi}A^{(-)}_\mu(+\infty)=-{N_c\over 2\pi}\,\,\partial_\mu^{(-)}\varphi_+\quad,
\ee
is its expectation value. Similarly, in terms of $\varphi_-\equiv \alpha(w=-\infty)$, the $A_\mu^{(-)}(-\infty)=\partial_\mu^{(-)} \varphi_-$ 
is a source that couples to the current $j_\mu^{(+)}$, and $-{N_c\over 2\pi}\,\,\partial_\mu^{(+)}\varphi_-$ is its expectation value.
Let's first see how the sourceless theory gives us the sought-for massless state in the spectrum.

Since $A^{(+)}_\mu(+\infty)=\partial_\mu^{(+)} \varphi_+$ is a source, it is turned off for dynamical modes. Thus the constraint
\be
\partial_\mu^{(+)}\varphi_+=0\quad,
\ee
has to be put and this yields $\varphi_+$ massless and chiral. The $j_\mu^{(-)}$ current is then given by $-{N_c\over 2\pi}\partial_\mu^{(-)}\varphi_+$.
Note that there is no action for $\varphi_+$, and its masslessness arises from the constraint, not
from the equation of motion. In some sense, there are no off-shell $p^2\neq0$ modes of $\varphi_+$ from the onset.
The same argument leads the constraint $\partial_\mu^{(-)} \varphi_-=0$. Then $\varphi_-$ is an oppositely chiral massless scalar and the $j_\mu^{(+)}$ current is given in terms of it as $-{N_c\over 2\pi}\partial_\mu^{(+)}\varphi_-$.

Although one might introduce 
\be
\varphi\equiv \varphi_+ - \varphi_-\quad,  \quad\tilde \varphi \equiv -\varphi_+ - \varphi_-\quad,
\ee
so that the vector/axial currents $j_{V,A}=j^{(+)}\pm j^{(-)}$ are written as
\be
j^\mu_V ={N_c\over 2\pi}\epsilon^{\mu\nu}\partial_\nu\varphi = {N_c\over 2\pi}\partial^\mu \tilde\varphi\quad, \quad
j^\mu_A = {N_c\over 2\pi}\partial^\mu \varphi={N_c\over 2\pi}\epsilon^{\mu\nu}\partial_\nu\tilde \varphi \quad,  
\ee
There is no difference between the two choices of $\varphi$ and $\tilde\varphi$ unless one switches on an external gauge potential 
that couples to them and insists on gauge invariance. Note that the physical degrees of freedom come only from the boundary values of the gauge 
functions $\alpha(\pm\infty)=\varphi_\pm$. The bulk gauge transformations that do not change the boundary values $\alpha(w=\pm\infty)=\varphi_\pm$ are unphysical and should be considered spurious. This is in complete analogy with~\cite{Sakai:2004cn}.

Let us extend the above discussion by introducing external vector gauge potential $V_\mu$ that couples to $j_V^\mu$. The example would be an electromagnetism of photon fields. This is an important testing ground  for the massless chiral bosons in holography. Indeed, It is well known 
that the photon gets massive once we make the photon field dynamical by introducing a kinetic term $-{1\over 4 e^2} F_{\mu\nu}^V F^{\mu\nu}_V$ for $V_\mu$ (Schwinger mechanism). The mass of the photon in our case is expected to be 
\be
m^2_\gamma= {N_c e^2 \over \pi}\quad.
\ee
We now show that this is indeed the case thanks to the 3D Chern-Simons term in the holographic dual.

Since $V_\mu$ acts as a source to the chiral currents and couples through
\be
V_\mu j^\mu_V= V_\mu^{(+)}j^{(-)\mu}+V_\mu^{(-)}j^{(+)\mu}\quad,
\ee
we now have the modified constraint equations for $\varphi_\pm$
\be
\partial_\mu^{(+)} \varphi_+ = V_\mu^{(+)}\quad,\quad 
\partial_\mu^{(-)} \varphi_- = V_\mu^{(-)}\quad.\label{modconst}
\ee
In bulk these are pure gauge functions, so their bulk action contribution is zero modulo the non-vanishing boundary counter terms
in the presence of the external source $V_\mu$. In particular,  the first term (\ref{1counter}) which is an important ingredient in the 3D 
holography of 3D the Chern-Simons term gives  the on-shell action
\be
{\cal L} = -{N_c\over 4\pi} V_\mu^{(+)}\partial^{(-)\mu}\varphi_+ 
-{N_c\over 4\pi} V_\mu^{(-)}\partial^{(+)\mu}\varphi_- \quad.
\label{phiac}
\ee
We have added the two contributions from both boundaries $w\to\pm\infty$. The other counterterms do not contribute 
in the limit $\epsilon\to 0$. 

The action (\ref{phiac}) is in fact not complete in view of gauge invariance of $V_\mu$, and should be supplemented with the Bardeen counterterm \cite{Bardeen:1969md}. 
One way to derive the Bardeen term is to look at the chiral current equations,
\bear
\partial_\mu j^{(+)\mu}&=&-{N_c\over 2\pi} \partial_\mu^{(-)} \partial^{(+)\mu}\varphi_- =
-{N_c\over 2\pi}  \partial^{(+)\mu}\partial_\mu^{(-)}\varphi_-=
-{N_c\over2\pi} \partial^{(+)\mu}V_\mu^{(-)}\quad,\nonumber\\
\partial_\mu j^{(-)\mu}&=&-{N_c\over 2\pi} \partial_\mu^{(+)} \partial^{(-)\mu}\varphi_+ =
-{N_c\over 2\pi}  \partial^{(-)\mu}\partial_\mu^{(+)}\varphi_+=
-{N_c\over2\pi} \partial^{(-)\mu}V_\mu^{(+)}\quad,
\eear
where we have used the constraint (\ref{modconst}) in the last equalities.
The vector current $j_V^\mu=j^{(+)\mu}+j^{(-)\mu}$ is not conserved, 
\be
\partial_\mu j_V^\mu = - {N_c\over 2\pi}\left(\partial^{(+)\mu}V_\mu^{(-)}+\partial^{(-)\mu}V_\mu^{(+)}\right) = -{N_c\over 2\pi}\partial_\mu V^\mu\quad.
\ee
Strict conservation of the vector current follows by adding the Bardeen term to the effective action,
\be
{\cal L}_{\rm Bardeen} = {N_c\over 2\pi} V_\mu^{(+)}V^{(-)\mu} = {N_c\over 4\pi} V_\mu V^\mu\quad,\label{bardeen}
\ee
so that the new currents are
\be
j^{(+)\mu}_{\rm new}= -{N_c \over 2\pi}\partial^{(+)\mu}\varphi_- +{N_c\over 2\pi} V^{(+)\mu}\quad,\quad
j^{(-)\mu}_{\rm new}= -{N_c \over 2\pi}\partial^{(-)\mu}\varphi_+ +{N_c\over 2\pi} V^{(-)\mu}\quad,
\ee
from which one easily checks that
\be
\partial_\mu j_V^\mu = 0\quad,\quad \partial_\mu j_A^\mu= {N_c\over 2\pi}\epsilon^{\mu\nu}F^V_{\mu\nu}\quad,
\ee
which exhibits the correct anomaly structure.

Another way of deriving the same Bardeen term is to require vector gauge invariance on the action (\ref{phiac}).
Under the transformation
\be
V_\mu \to V_\mu + \partial_\mu \alpha\quad,
\ee
the constraints (\ref{modconst}) imply that one also needs to transform $\varphi_\pm \to \varphi_\pm+\alpha$ for consistency.
With these, the action (\ref{phiac}) transforms as
\be
\delta {\cal L}= -{N_c\over 4\pi}\left(\partial^{(+)}_\mu \alpha \partial^{(-)\mu}\varphi_+  
+\partial^{(-)}_\mu \alpha \partial^{(+)\mu}\varphi_- +V_\mu^{(+)}\partial^{(-)\mu}\alpha 
+V_\mu^{(-)}\partial^{(+)\mu}\alpha\right) = -{N_c\over 2\pi} V_\mu \partial^\mu\alpha \quad,
\ee 
where we performed partial integrations and used the constraints (\ref{modconst}).
The Bardeen term (\ref{bardeen}) has precisely the opposite variation $\delta {\cal L}_{\rm Bardeen} = {N_c\over 2\pi} V_\mu \partial^\mu\alpha$
to ensure the invariance of the total action.

With this in mind, the vector conserving effective action then reads as
\be
{\cal L}_{\rm eff} = -{N_c\over 4\pi} V_\mu^{(+)}\partial^{(-)\mu}\varphi_+ 
-{N_c\over 4\pi} V_\mu^{(-)}\partial^{(+)\mu}\varphi_- +{N_c\over 2\pi} V_\mu^{(+)}V^{(-)\mu} \quad,
\ee
with the constraints (\ref{modconst}): $\partial_\mu^{(+)} \varphi_+ = V_\mu^{(+)}$, $\partial_\mu^{(-)} \varphi_- = V_\mu^{(-)}$.
Starting from this, we now show that upon making $V_\mu$ dynamical one gets to a theory of massive photon corresponding to the Schwinger mechanism.
One remark is that the above is derived from the pure gauge sector with the boundary counterterm (\ref{1counter}), so it is equally true
in the case of pure 3D Chern-Simons theory which is dual to 2D chiral boson theory as it should be. 

The gauge transformation of $\varphi_\pm$ makes it natural to reparametrize them as
\be
\varphi_\pm = \alpha \pm \varphi\quad,
\ee
with a slight abuse of notation; here $\alpha$ and $\varphi$ are newly introduced dynamical fields.
The action in terms of these new fields reads
\bear
{\cal L}_{\rm eff} &=& -{N_c\over 4\pi} V_\mu \partial^\mu\alpha +{N_c\over 4\pi}V_\mu \,\epsilon^{\mu\nu}\partial_\nu \varphi
+{N_c\over 4\pi} V_\mu V^\mu \nonumber\\
&-&{1\over 4 e^2} F_{\mu\nu}^V F^{\mu\nu}_V +\eta_\mu \left(V^\mu -\partial^\mu\alpha -\epsilon^{\mu\nu}\partial_\nu\varphi\right)\quad,\label{action1}
\eear
where in the second line we included the kinetic term for the external  dynamical $V_\mu$.
Also we  imposed the constraints (\ref{modconst})
explicitly in the action by introducing a Lagrange multiplier $\eta_\mu$.  
The above action (\ref{action1}) is gauge invariant up to surface terms under 
\be
V_\mu \to V_\mu+\partial_\mu \Lambda\quad,\quad \alpha \to\alpha+\Lambda\quad,\quad \eta_\mu \to\eta_\mu -{N_c\over 4\pi}\partial_\mu\Lambda\quad,
\ee
and one has to fix the gauge to discuss physical degrees of freedom.
One convenient gauge fixing is the analog of the $R_\xi$-gauge,
\be
{\cal L}_{\rm gauge\,\,fixing}= -{1\over 2e^2 \xi} \left(\partial_\mu V^\mu + \xi {N_c e^2\over 4\pi} \alpha\right)^2\quad,
\ee
that removes mixing between $V_\mu$ and $\alpha$. The resulting gauge-fixed action is algebraically quadratic in $\alpha$  and after
integrating out $\alpha$, one arrives at
\bear
{\cal L}_{\rm eff} &=& -{1\over 4 e^2} F_{\mu\nu}^V F^{\mu\nu}_V 
+{N_c\over 4\pi} V_\mu V^\mu-{1\over 2 e^2\xi}\left(\partial_\mu V^\mu\right)^2 +{N_c\over 4\pi}V_\mu \,\epsilon^{\mu\nu}\partial_\nu \varphi \nonumber\\
&+&\eta_\mu \left(V^\mu -\epsilon^{\mu\nu}\partial_\nu\varphi\right) + {1\over 2 e^2\xi}\left(4\pi\over N_c\right)^2 \left(\partial_\mu\eta^\mu\right)^2 \quad.\label{action2}
\eear
Although the theory should be independent of $\xi$, it is most convenient to consider the unitary gauge $\xi\to\infty$ upon which one simply removes the last term in (\ref{action2}). Then, integrating over $\eta_\mu$ gives us the constraint
\be
V^\mu =\epsilon^{\mu\nu}\partial_\nu\varphi\quad,\label{const3}
\ee
which is mathematically equivalent to $\partial_\mu V^\mu=0$ in two dimensions. The final action then becomes
\be
{\cal L}_{\rm eff}= -{1\over 4 e^2} F_{\mu\nu}^V F^{\mu\nu}_V 
+{N_c\over 4\pi} V_\mu V^\mu+{N_c\over 4\pi}V_\mu \,\epsilon^{\mu\nu}\partial_\nu \varphi=
-{1\over 4 e^2} F_{\mu\nu}^V F^{\mu\nu}_V 
+{N_c\over 2\pi} V_\mu V^\mu\quad,
\ee
where we used the constraint (\ref{const3}) in the last equality. Combined with (\ref{const3}), $\partial_\mu V^\mu=0$,
this is precisely the theory of a massive photon in the axial gauge $\partial_\mu V^\mu=0$ with the mass
\be
m_\gamma^2={N_c e^2\over\pi}\quad,
\ee
in agreement with the Schwinger mechanism.

\section{Conclusions}

We have presented a holographic formulation of 1+1 massless QCD based on a $D2/D8/\overline{D8}$ construction
that is very similar to the holographic formulation of 1+3 massless QCD~\cite{Sakai:2004cn}. While this construction
was visited before in~\cite{Gao:2006up,Rodriguez:2005jr}, the leading role played by the 3D Chern-Simons term
was overlooked. The emergence of a massless chiral boson, and a normalizable heavy meson spectrum relies heavily
on the 3D Chern-Simons term. 

Unlike in 1+3 dimensions, the massless chiral boson completly decouples from the massive chiral mesons, which is a particular
feature of the chiral anomaly. Also, all two-point functions are saturated by the massless chiral boson in complete
agreement with the resummation of the planar graphs at large $N_c$~\cite{GROSS}.  These features will be relaxed
by the introduction of masses for the quarks. Massless 1+1 QCD at large $N_c$ exhibits anti-vector dominance.

In the presence of dynamical photons, we have shown that the massless chiral boson yields a photon mass in 
complete agreement with the Schwinger mechanism. This vindicates completely the identification of the holographic
chiral and massless mode with the boundary holonomy.

The $D2/D8/\overline{D8}$ embedding supports instanton-like flux configurations in the bulk for the $S^6$-wrapped $D6$ branes, which are likely the precursors of the chiral
baryons at the boundary. We hope to address the issues related to the baryon and massive meson spectrum for a
finite current quark mass in the near future.

\vskip 1cm \centerline{\large \bf Acknowledgement} \vskip 0.5cm

We thank Hans Hansson, Dima Kharzeev and Toru Kojo for helpful discussions.
This work was supported in part by the U.S. Department of Energy under Contract No.~DE-FG02-88ER40388.

 \vfil

\end{document}